\documentstyle[12pt]{article}
\makeatletter
\if@twoside
\else \oddsidemargin 00pt \evensidemargin 00pt
\fi
\let\@eqnsel = \hfil

\expandafter\ifx\csname mathrm\endcsname\relax\def\mathrm#1{{\rm #1}}\fi
\@ifundefined{mathrm}{\def\mathrm#1{{\rm #1}}}{\relax}

\makeatother

\def\gtrsim{\ \rlap{\raise 2pt \hbox{$>$}}{\lower 2pt \hbox{$\sim$}}\ }
\def\lesssim{\ \rlap{\raise 2pt \hbox{$<$}}{\lower 2pt \hbox{$\sim$}}\ }

\def\ea{{et al.}}
\def\ib{{\it ibid.}}

\def\npb#1{Nucl. Phys. {\bf B #1}}
\def\plb#1{Phys. Lett. {\bf B #1}}
\def\prd#1{Phys. Rev. {\bf D #1}}
\def\prl#1{Phys. Rev. Lett. {\bf #1}}
\def\ncim#1{Nuo. Cim. {\bf #1}}

\def\apj#1{Astrophys.\ J. {\bf #1}}
\def\nature#1{Nature {\bf #1}}

\def\hb{\hfill\break}

\begin{document}
\thispagestyle{empty}
\null
\vskip -1truecm
\hfill UM-TH-94-36

\hfill INFN-FE-06-94

\hfill hep-ph/9411249

\hfill October 1994

\vskip .5cm

\begin{center}
{\Large \bf      %INSERT TITLE
NATURAL R PARITY CONSERVATION \\ WITH HORIZONTAL SYMMETRIES. \\
A FOUR GENERATION MODEL
\par} \vskip 2.em

{\large		%INSERT NAMES
{{\sc Zurab Berezhiani}$^{a,b}$ and {\sc Enrico Nardi}$^{c}$}
\\[1ex] %INSERT ADDRESS
{$^{a}$\it INFN, Sezione di Ferrara, 44100 Ferrara, Italy}\\
%{and}\\
{$^{a}$\it Institute of Physics, Georgian Academy of Sciences, 380077
Tbilisi, Georgia}\\
{$^{c}$\it Department of Particle Physics, Weizmann Institute of Science,}\\
{  P.O.B. 26, Rehovot, 76100 Israel}\\ [2ex]
\par}
\end{center} \par
\vskip .5truecm
\noindent
{\bf Abstract}
\par\noindent
In most supersymmetric models the stability of the proton is ensured
by invoking R-parity. A necessary ingredient to enforce R-parity is
the possibility of distinguishing the lepton  superfields from
the Higgs ones. This is generally achieved either by assuming
different charges under some matter parity, or by assigning the
superfields to different representations of a unified gauge group.
We want to put forward the idea that the replica of the fermion
generations, which constitute an intrinsic difference between
the fermions and the Higgs superfields, can give a clue to understand
R-parity as an accidental symmetry. More ambitiously, we suggest a possible
relation between proton stability and the actual number of fermion generations.
We carry out our investigation in the framework of non-Abelian horizontal gauge
symmetries. We identify $SU(4)_H$ as the only acceptable horizontal gauge
 group whichcan naturally ensure the absence of R parity violating operators,
without conflicting with other theoretical and phenomenological constraints.
We analyze a version of the supersymmetric standard model equipped with a
gauged horizontal $SU(4)_H$, in which R-parity is accidental.  The model
predicts four  families of fermions, it allows for the dynamical generation of
a realistic hierarchy of fermion masseswithout any ad hoc choice of small
Yukawa couplings,  it ensures in a natural way the heaviness of all
the fourth family fermions (including the neutrino) and it predicts a {\it
lower} limit for the $\tau$-neutrino mass of a few eV. The scale of the
breaking of  the horizontal symmetry can be constrained rather precisely
in a narrow window around $\sim 10^{11}$ GeV. Some interesting astrophysical
and cosmological implications of the model are addressed as well.
\par
\vskip .5cm
\noindent
%PACS number(s):
\noindent
--------------------------------------------\phantom{-} \hb
% \leftline{$\dagger$ Address from November 1994: {\it Department of Particle
%%Physics, }}
% \leftline{{\it  Weizmann Institute of Science,}  P.O.B. 26, Rehovot, 76100
%%Israel}
\vskip .1mm
\noindent
Electronic mail:
berezhiani@fe.infn.it,
ftnardi@wiswic.weizmann.ac.il
%\vskip .5cm
%\noindent UM-TH-94-06-?? \par
%\noindent INFN-FE-06-94 \par
%\vskip .15mm
%\noindent October 1994 \par
%\vskip 1cm
\null
\setcounter{page}{0}
\clearpage

\section{Introduction}

In the Standard Model (SM), Baryon (B) and Lepton (L) numbers are
conserved as a result of accidental global $U(1)_B$ and $U(1)_L$
symmetries that follow from the requirement of gauge invariance and
renormalizability.
These symmetries are violated only by higher order non-renormalizable
operators, cutoff at the Planck scale, which can arise from
non-perturbative quantum gravity or string effects \cite{non-ren-op}.
\footnote{
For example, the dimension 5
lepton number violating term $(1/M_{Pl})ll\Phi\Phi$
provides a neutrino Majorana mass of about $10^{-5}$ eV, which
could be relevant for the solar neutrino oscillations.
However, analogous terms violating baryon number are
dimension 6 or higher, and hence too
small to cause any observable effect. }
In the Supersymmetric (SUSY) version of the Standard Model (SSM) this
is not true anymore.
Consider in fact the quark and lepton left-handed chiral superfields,
which transform under SU(3) $\times$ SU(2) $\times$ U(1) as follows:
\begin{eqnarray}
&q = {u\choose d} \sim (3,2,1/6),~~~~~~
u^c \sim (\overline 3, 1, -2/3),~~~~~~
d^c \sim (\overline 3, 1, 1/3), \cr
&l = {\nu\choose  e} \sim (1,2,-1/2),~~~~~~
e^c \sim (1,1,1).
\label{fields}
\end{eqnarray}
The two Higgs superfields $\Phi_{1,2}$ transform as $(1,2,\mp 1/2)$
respectively.
As the scalar components of $\Phi_{1,2}$ acquire
nonzero vacuum expectation values (VEVs), the fermions acquire a mass
through the superpotential terms
$\Phi_1 q d^c$, $\Phi_2 q u^c$ and $\Phi_1 l e^c$.
As for the Higgsinos, their mass is provided by the SUSY invariant
term $\mu \Phi_1 \Phi_2$.
Since $l$ and $\Phi_1$ have the same transformation properties
under the gauge group, the L and B violating terms obtained by
substituting $\Phi_1 \to l$ are also allowed by the gauge symmetry,
as well as an additional
term involving three quark superfields. These terms
read (family and gauge indices are suppressed)
\begin{equation}
\mu^\prime\, l \Phi_2, \quad \lambda\, l l e^c, \quad
\lambda^\prime\, l q d^c, \quad  \lambda^{\prime\prime}\, u^c d^c d^c.
\label{Rviol}
\end{equation}
The simultaneous presence of all the terms in
(\ref{Rviol}) is phenomenologically unacceptable.
In particular, if both the third and fourth of these terms are present, their
combination would lead to catastrophically fast proton decay
mediated by $d^c$-type squark exchange, unless the relevant
couplings are fine-tuned to extremely small values
$\lambda^\prime \cdot \lambda^{\prime \prime}
\lesssim 10^{-26} (\frac{m_{\tilde{d^c}}}{1 {\rm  TeV}})^2$.
As for the $\mu^\prime$  term, in the SSM it is possible to eliminate
it
through a suitable rotation among the $l$ and $\Phi_1$ superfields
\cite{Hall-Suzuki}. After such a redefinition, the mass Yukawa terms
$\Phi_1 q d^c$ and $\Phi_1 l e^c$ give rise to (corrections to)
the $\lambda$ and $\lambda^\prime$ terms in  (\ref{Rviol}).
However, in
some extensions of the SSM $l$ and $\Phi_1$ will not have the same
quantum numbers, thus this rotation is not always possible so that we will keep
explicitely the $\mu^\prime$ term.\footnote{This term, together with
an analogous term in the scalar potential which leads to
 non-vanishing sneutrino VEVs, induces
a mixing between the neutrinos and the neutralinos,
and can easily generate a too large value for the $\nu$ mass.}

The relevant symmetry that ensures the B and L conservation in the SSM
is called R parity, which is defined as $\rm R \equiv (-1)^{2J+3B+L}$,\
where J is the spin of the particle and  B(L) its baryon (lepton) number
\cite{R-parity}. R parity
is an automatic consequence of a $Z_2$ matter
parity under which the fermion superfields change the sign while the
`Higgs' ones $\Phi_{1,2}$ remain invariant.
R parity is trivially related to $Z_2$ matter parity
by a factor of -1 for fermions, and hence
does not commute with supersymmetry.
It is a well known fact that an unsatisfactory feature of the
SSM is that the $Z_2$ (or equivalently R) parity
conservation has to be imposed by hand.

In the context of Grand Unification Theories (GUT) based
on the gauge group $SU(5)$, the fermion superfields are assigned
to the $\bf 10 +\bar {\bf 5}$ representation of $SU(5)$,
while $\Phi_1$ and $\Phi_2$ belong respectively to the
$\bar {\bf 5}_{\Phi_1}$ and ${\bf 5}_{\Phi_2}$.
The down-quark and lepton masses are generated through the coupling
${\bf 10} \, \bar {\bf 5}\, \bar {\bf 5}_{\Phi_1}$,
and the gauge invariant terms obtained by
$\bar {\bf 5}_{\Phi_1}\to \bar {\bf 5}$, (namely
${\bf 10} \, \bar {\bf 5}\, \bar {\bf 5}$ and
and $\bar {\bf 5}\, {\bf 5}_{\Phi_2}$)
lead again to the set of B and L violating couplings in (\ref{Rviol}).
Thus, with respect to automatic R-parity conservation
the supersymmetric  $SU(5)$ model does not differ much from the SSM.
In addition to this, in SUSY $SU(5)$ the effectiveness of imposing
 R parity  as an additional
global discrete symmetry is also questionable. In fact,
 even if at the renormalizable level the  ${\bf 10 \bar5\bar5 }$  term is
forbidden,
 a  problem can appear due
to non-perturbative quantum gravity effects (virtual black holes or wormholes)
 which in general do not respect the global charges.
 These effects can induce in the superpotential  higher order terms as
$(1/M_{Pl}) {\bf \bar 5\, \bar 5\, 10\, 24 }$, where the $\bf 24$ is
the adjoint  Higgs representation which breaks the  $SU(5)$ symmetry. After
substituting its VEV ($\langle {\bf 24} \rangle \sim 10^{16}$ GeV), this
operator
reduces to the terms given in (2) with a Yukawa coupling  $\langle {\bf 24}
\rangle
/M_{Pl} \sim 10^{-3}$, which would be catastrophic for proton decay.
This argument supports the idea that
it is highly desirable to achieve R parity conservation in an automatic
way as a natural consequence of gauge invariance, which could then  protect the
symmetry
from effects of this kind.

As is well known, this is the case for  $SO(10)$ models. Indeed
$SO(10)$ offers an elegant solution to this problem,
since the fermion superfields are in the spinor representation {\bf 16}
whereas the Higgs ones are generally assigned to vector
representations as {\bf 10}, {\bf 45}, {\bf 54}, {\bf 126} etc..
%assigned to different representations
%(respectively the {\bf 10}$_\Phi$ and ).
The masses of the fermions, including the
neutrinos, can be generated through the gauge invariant couplings
${\bf 16}\, {\bf 16}\, {\bf 10}_\Phi$ and ${\bf 16}\, {\bf 16}\,
{\bf 126}_\Phi$ \cite{Mohapatra}, while the terms
% ${\bf 16}\, {\bf 16}\, {\bf 16}$
$({\bf 16})^3$ and ${\bf 16}\, {\bf 10}_\Phi$ are
 forbidden since they are not  $SO(10)$
invariant. In other words, as long as all the $SO(10)$ invariant couplings
allow for only pairs of 16-plets,
the theory has an automatic $Z_2$ matter parity
under which 16-plets change the sign whereas the superfields
in vector representations remain
invariant. This is not true anymore for the $SO(10)$
models in which the symmetry breaking is triggered also
by the scalar components of superfields
belonging to the ${\bf 16}_{\Phi}+{\bf \overline{16}}_{\Phi}$.
(Examples of SUSY models in which these representations
play the role of the standard
${\bf 126}+{\bf \overline{126}}$ can be found in ref.
\cite{BabuBarr}).
In fact, after substituting the VEV $\langle {\bf 16}_{\Phi} \rangle$,
the couplings $\frac{1}{M}{\bf 16}^3{\bf 16}_{\Phi}$ which are allowed by
the gauge symmetry lead again to R-parity violating terms.
Since in these models the right handed neutrino masses are generated by
operators $\sim \frac{1}{M}{\bf 16}^2 {\bf 16}^2_{\Phi}$
which have the same structure,
the ratio $\langle {\bf 16}_{\Phi} \rangle /M $ cannot be very
small, implying that the magnitude of the resulting R-parity
violating terms is again in conflict with the limits on the proton
lifetime. In addition to this, a direct term
$m^2 {\bf 16} {\bf 16}_{\Phi}$ in the scalar potential
will induce a nonvanishing VEV for some scalar field carrying lepton
number, leading also to spontaneous R-parity violation trough
sneutrino VEVs.
We conclude that R-parity conservation is not automatic
anymore for the $SO(10)$ models with Higgs fields belonging to the
${\bf 16}_\Phi$. In this case some additional discrete
symmetry has to be imposed by hand in order to distinguish the
fermion 16-plets from the Higgs ones.

The result of this brief analysis shows that in order to ensure
the absence of the operators (\ref{Rviol}),
some `label'  to  distinguish  the Higgs superfields
from the fermion ones is required.
Such a label can be chosen {\it ad hoc}, as in the SSM and in the $SU(5)$
model, by assigning
to the Higgs and fermion superfields different
$Z_2$ parities or, as in $SO(10)$, it can arise
in a more natural way by
assigning the different superfields
 to different representations of the GUT gauge group.

Our work stems from the observation that
there is a natural  distinction between
the fermion and the Higgs superfields: namely that {\it fermions
superfields replicate in different generations, while Higgs
superfields do not}.
In this paper we wish to investigate if, in the framework of
SUSY models, any natural link can be found between two
striking but apparently unrelated experimental evidences:
the stability of the proton (or more generally the conservation of
R-parity) and the replica of fermion generations.
More precisely, we want to put forward the idea that
R (or equivalently $Z_2$) parity  could arise
as a consequence of a {\it gauged} horizontal symmetry,
broken at some high scale,
which constitutes also  a natural framework to account for the
replica of  fermion generations.
Clearly,
the horizontal group $G_H$ should act only on the quark-lepton superfields,
while the Higgses $\Phi_{1,2}$ responsible for the electroweak
symmetry breaking are $G_H$-singlets. In such a picture,
 independently of the choice of the vertical gauge group and/or
of the particular superfield assignments to its representations,
the Higgs and fermion superfields can be always distinguished
by their different transformation properties under the
 horizontal gauge group, and
 this leads to the possibility of allowing the necessary mass
terms which are bilinear in the fermion superfields, while
forbidding the B and L violating linear and
trilinear couplings in (\ref{Rviol}).

Beyond accounting for the number of generations,
models based on gauged horizontal symmetries
 present several additional interesting features
(see for example the models
\cite{horizontal,Khlopov} based on $SU(3)_H$ horizontal symmetry).
They naturally embed the  % esthetical
principle of `flavor democracy'
\cite{fl-democracy}, namely that all fermions with the same
gauge quantum numbers should have the same short distance
interactions, while the observed mass differences arise
from dynamics at some large energy scale.
They can also explain qualitatively the observed pattern
of fermion masses and mixing  without appealing to unnaturally small
values for any fundamental parameter \cite{horizontal,Khlopov}.
The structure of the fermion mass matrices can in fact
be related to the horizontal
symmetry breaking pattern. Then the
mass hierarchy between families
arises dynamically from certain hierarchies
in this breaking, while the fermion Yukawa
couplings are generation blind, and can be assumed
to be all of order unity.

The paper is organised as follows: in Section 2 we will carry
out a general analysis of the non-Abelian horizontal
symmetries that can forbid the R-parity violating operators
(\ref{Rviol}), and that at the same time satisfy a certain number of
theoretical and phenomenological constraints. We will show
that $SU(4)_H$, with the fermions assigned to the
fundamental {\bf 4}$_H$ representation, is the only viable candidate.
Then in our scheme the presence of one additional family of fermions
results as  an unavoidable prediction.
In Section 3 we will explicitly construct a model based on
$SU(4)_H$. We will discuss the pattern of the Horizontal symmetry
breaking, and the peculiar form of the resulting
fermion mass matrices. We will show that the minimal number
of horizontal scalars needed to reduce completely the rank of the
group, can
also ensures that all the fourth family
fermions (including the neutrino)  acquire naturally large masses, typically of
the order of
 the electroweak scale. An additional interesting feature of our model is that
the fourth family fermions
 are unmixed with the lighter ones.
In Section 4 we will address some of the
phenomenological consequences of the model.
With the aid of cosmological and astrophysical arguments,
the scale at which the horizontal symmetry is completely broken
can be constrained to a narrow window around $10^{11}$ GeV.
This results in a strict {\it lower} limit of a few eV
for the mass of the $\tau$ neutrino.
We also discuss the possible decay channels and lifetime for
the unmixed fourth family quarks. We show how our model
leads to the prediction of
a possible cosmological signal of diffuse gamma ray
flux over background from decays of relic $b^\prime$.
The problem of the baryogenesis mechanisms viable
in our model is also briefly addressed in this section.
Finally, in Section 5 we will collect the main results and draw
our conclusions.

\section{General Analysis of Horizontal Gauge Groups}

Our aim is to find and classify the theories in which the
horizontal gauge group $G_H$ naturally forbids the terms in
(\ref{Rviol}) due to gauge principles, or in other words in which
R parity (or equivalently $Z_2$ matter parity) appears as an
automatic consequence of the horizontal gauge symmetry and of the
field content of the model. We demand that the models we are
interested in should satisfy the following list of
basic requirements:

$(i)$ In order to ensure a straightforward definition for the
horizontal gauge symmetry as a symmetry intrinsically related
to the number of fermion generations,
all the fermion superfields with the same quantum numbers
 must  fill up  one irreducible
representation of the horizontal group $G_H$. In particular, we
forbid $G_H$ singlet families.

% For example, for the group $G_H=SU(N)$ each fermion superfield can be
% assigned either to the $N$ or to the $\overline N$ representation of
% the group.

$(ii)$
On the other hand, in order to implement consistently our idea,
the standard Higgs superfields $\Phi_{1,2}$ must be singlets with respect
to $G_H$.
This requirement is needed also to prevent the proliferation of Higgs
doublets with masses at the electroweak scale, that would spoil the
natural suppression of  Flavor Changing Neutral Currents
(FCNC)  \cite{FCNC}. Moreover, Higgs in non-singlet
horizontal representations would also destroy gauge coupling
unification, thus preventing any attempt to embed the model in some
vertical GUTs.

$(iii)$ We demand that the couplings in (\ref{Rviol}) are forbidden
as a consequence of conditions $(i)$ and $(ii)$,
together with the requirement of horizontal gauge invariance.
Clearly, any non-Abelian group satisfying  $(i)$ and  $(ii)$
will forbid the first term in  (\ref{Rviol}), since it is linear in the
fermion superfields, so in the following we will concentrate on the
additional conditions needed to forbid the trilinear terms.

Since we are investigating the possibility of relating
the absence of R-parity violating operators to the number of generations,
we wish to treat the latter one as a free parameter, to be determined
by the dimension $N$ of the representation of the
non-Abelian horizontal group. However, in order
to have phenomenologically realistic theories,  $N$
should not be too large.
For example additional families would contribute to
the radiative corrections to electroweak
quantities. Detailed analyses of the precise electroweak data
have been recently performed, and they rule out $N\geq 6$
\cite{fourth-fam}.  We will then restrict our analysis to groups with
representations of dimension $N = 3,\, 4,\, 5 $.

To have phenomenologically appealing models, the
following additional constraints should be also imposed:

$(iv)$
We require that a realistic pattern of fermion masses and mixings
should arise {\it naturally} as a result of the dynamics of the horizontal
symmetry breaking, and in particular, since no new state in addition to the
three families of fermions have been observed yet,
possible new generations should be naturally heavy.

More specifically, the fermion masses should arise
from effective operators with the structure
\begin{equation}
{\cal O}_{\rm eff} \sim \frac{{\cal P}^{(n)}(\xi_k)}{M^n}\, f f^c
\Phi_{1,2},
\label{mass-op}
\end{equation}
where $f,f^c$ are the fermion superfields in eq. (\ref{fields}),
${\cal P}^{(n)}$ represents some $n$-order polynomial of the
scalars $\xi_k$ responsible for the breaking of $G_H$,
and $M$ is some cutoff mass scale.\footnote{ The non-renormalizable
couplings (\ref{mass-op}) with $M\sim M_{Pl}$ could appear due to
quantum gravity effects. Alternatively, these operators with arbitrary
$M$ can be effectively generated through the exchange of some superheavy
fields with $O(M)$ masses \cite{Frogatt}. }
Clearly, in order to ensure $G_H$
invariance,  ${\cal P}^{(n)}$ should transform as the conjugate
of the tensor product of $f$ and $f^c$.

This `naturalness' condition is a rather strong one,
since it rules out at once all the self-conjugate representations
${\cal N}_S$, as well as the groups that have only
self-conjugate representations.
In fact, the horizontal gauge invariant term
$ {\cal N}_S\, {\cal N}_S\, \Phi_{1,2}$
would imply equal masses for the different generations.
The mass splitting between different families could then be achieved
by means of the additional effective operators (\ref{mass-op})
only in a very unnatural way, at the price of {\it many}
fine tunings or  {\it ad hoc} choices for the relevant parameters.
Indeed,
 generating a hierarchy in this way can be hardly regarded as realistic.

$(v)$ An additional  condition is that R-parity breaking terms
should not appear even after $G_H$ breaking. More precisely,
all effective operators of the form
\begin{equation}
\frac{{\cal P}^{(n)}(\xi_k)}{M^n}\, f f f^c, \quad
% \frac{{\cal P}^{(n)}(\xi_i)}{M^n}\, l l e^c, \quad
\frac{{\cal P}^{(n)}(\xi_k)}{M^n}\,  f^c f^c f^c,
\label{Rvioloper}
\end{equation}
should be forbidden by the $G_H$ symmetry, since after the horizontal
symmetry breaking $\xi_k \rightarrow \langle\xi_k\rangle$ these terms
would generate again the R parity violating couplings (\ref{Rviol}).

\vskip .3truemm

We start our analysis with the only simple groups that
have three dimensional representations, namely
$SO(3)_H$ and $SU(3)_H$.

For $SO(3)_H$, the term {\bf 3}$^3$ (as well as {\bf 5}$^3$)
contains a gauge singlet, and hence does not forbid
the couplings (\ref{Rviol}). Moreover,
$SO(3)_H$ contains only self-conjugate representations,
so even if we had to assign the fermions to the
{\bf 4}, the  phenomenological requirement $(iv)$
would not be satisfied.

For $SU(3)_H$ there are two possibilities.
{\it Vectorlike} $SU(3)_H$, with $q,l$
transforming as {\bf 3} and $u^c,d^c,e^c$ as ${\bf \overline 3}$,
is an interesting possibility, since it
forbids the second and third  terms in (\ref{Rviol}),
which is enough to ensure the proton stability. However,
the $SU(3)_H$ invariant terms $f f^c \Phi_{1,2}$ is allowed,
and thus once more condition $(iv)$ is not fulfilled.
In addition vectorlike $SU(3)_H$ would also impede
the unification of the fermions (1) within one
irreducible GUT multiplet.

{\it Chiral} $SU(3)_H$, with all the fermions assigned to the
same representation {\bf 3} (or ${\bf \bar 3}$)
fails to satisfy condition $(iii)$, since
{\bf 3}$^3$ contains a gauge singlet.
However, models based on chiral $SU(3)_H$ \cite{horizontal,Khlopov}
have proven to be quite effective in relating the fermion mass
hierarchy to the hierarchy in the $SU(3)_H$ breaking, that is to the
hierarchy among the horizontal VEVs $\langle \xi_k\rangle $.
Since this success in accounting for
the pattern of fermion masses and mixings
is intimately related to the non self-conjugate nature
of the fundamental representation, it appears to be a general feature
of chiral $SU(N)$ models. This  points towards
chiral $SU(N)_H$ ($N > 3$) as possible interesting
candidates, since we immediately notice that in this case
the term $N^3$ does not contain gauge singlets,
and thus the terms in (\ref{Rviol}) are automatically forbidden.

As a first result, our analysis implies that in order to implement
our scheme the number of generations must be larger than 3,
and also suggests that the $SU(N)_H$ ($N > 3$)
groups represent a class of interesting candidates.

We now move to  groups with 4 dimensional
representations.
Apart from $SU(4)$, which we will discuss in detail in the following,
also for $SO(4)~~ (\sim SU(2)\times SU(2)) $ and $SO(5)~~ (\sim Sp(4))$
the lowest dimensional representation is the {\bf 4}.
In both cases, while the {\bf 4}$^3$ terms  are forbidden
and thus condition $(iii)$ is fulfilled,
the requirement $(iv)$ is not satisfied.
In fact, both these groups have only self-conjugate representations,
 the invariant Yukawa terms $f f^c \Phi_{1,2}$
are again allowed, and a hierarchy in the fermion masses cannot be
generated in a natural way.

Apart for $SU(5)$, there are no new groups with five dimensional
representations.As we will now show, the last
condition $(v)$ restrict the viable $SU(N)_H$
models to the cases when $N$ is even, thus ruling out
$SU(5)$ as a satisfactory horizontal symmetry.
Consider in fact $SU(N)_H$ with
the $f$ and $f^c$ fermion superfields assigned to the fundamental $N$
dimensional representation. The mass terms transform as $N\times N$ and
thus belong to two-index (symmetric and antisymmetric) representations.
In order to construct horizontal gauge invariant mass terms,
  we can take also
the horizontal  Higgses $\xi_k$  in two-index
representations. Then for $N$=4,6,\dots terms of the form
$N\times N \times N \times {\cal P}^{(n)}$
(that is the R-parity violating effective operators (\ref{Rvioloper}))
cannot arise, since it is impossible to saturate  all the indices
and construct horizontal gauge invariants.
In contrast, for $SU(N)_H$ with $N$ odd the totally antisymmetric
$\epsilon$ tensor allows to rewrite some combinations of Higgs fields
with an even number of free indices as tensors with an odd number of
free indices, which are suitable for generating gauge invariants when
matched with the $N\times N\times N$ term.
Indeed, after the horizontal symmetry breaking ($\xi\rightarrow
\langle \xi \rangle$) operators of the form
\begin{equation}
f^{(c)}_\alpha f^{(c)}_\beta f^{(c)}_\gamma
% \frac{\xi_1^{\delta\mu}\xi^2_{\mu\rho}\xi_3^{\rho\sigma}}
\frac{(\xi_1 \xi_2 \dots \xi_n)_{\delta\dots\sigma}}
{M^n}
\epsilon_{\alpha\beta\gamma \delta \dots\sigma}
\label{sun-odd}
\end{equation}
(which can be constructed also when the $\xi$'s
belong to the symmetric part of $N\times N$) will again
spoil R-parity.

Our analysis suggests that natural conservation of
R-parity, complemented with the additional phenomenological
constraints $(iv)$ and $(v)$, can
 be achieved in models based on chiral horizontal
symmetries $SU(N)_H$ with $N$ even, under which the quark and lepton
superfields transform as  fundamental $N$-plets.
The unwanted terms
transforming as $N^3$ are automatically forbidden by horizontal
gauge invariance, and the generation of a realistic pattern of
masses and mixings appears viable.
Clearly, in order to implement our scheme,
the number of families must be extended
to $N_f > 3$ and even.
As is well known, the possibility of extra families with a light
neutrino is ruled out by the results of the Mark II and LEP
collaborations \cite{three_nu}. However, these results do not exclude
sequential generations with
heavy  neutrinos ($m_\nu > M_Z/2$). On the other hand,
as we have already mentioned, detailed studies \cite{fourth-fam} of
the effects of radiative corrections due to additional families show
that precise electroweak data are not incompatible with a fourth
family, while six families (which would be our next interesting case)
are ruled out \cite{fourth-fam}. In addition to this, a dedicated analysis
showing the viability of supersymmetric models with four families
with respect to gauge coupling unification was presented in ref.
\cite{fourth-susy}. These results are relevant for our analysis,
since  condition $(ii)$ ensures that the field content
%of non singlet $SU(3)\times SU(2)\times U(1)$ fields
in our $SU(4)_H$ model is the same than that of the four family SSM
of ref. \cite{fourth-susy}, up to some large energy scale where
the horizontal symmetry breaks down (see Sect. 3).

We can conclude that the stability of the
proton in SUSY models can be naturally related to
the replica of the fermion generations by assuming a suitable
horizontal gauge symmetry. Such a symmetry ensures
that R parity arises as an accidental symmetry of the
gauge model. Moreover, theoretical and
phenomenological constraints allow to single out
$SU(4)_H$ as the only satisfactory horizontal gauge group,
on which we will concentrate in the rest of the paper.

\section{A Model with Horizontal Symmetry $SU(4)_H$}

Let us now consider the standard SU(3)$\times$SU(2)$\times$U(1)
vertical gauge group, with local chiral $SU(4)_H$ horizontal symmetry
acting on four families of left chiral superfields
\begin{eqnarray}
f_\alpha: \quad & \quad
q_\alpha& = {u \choose  d}_\alpha\sim (3,2,1/6,4), \quad\quad\quad
l_\alpha = {\nu \choose  e}_\alpha \sim (1,2,-1/2,4) \cr
f^c_\alpha: \quad &
u^c_\alpha& \sim (\bar{3},1,-2/3,4), ~~~
d^c_\alpha\sim (\bar{3},2,1/3,4),      \quad ~~
%\hskip 1truecm
e^c_\alpha \sim (1,1,1,4)
%\hskip 2.5truecm \alpha=1,2,3,4
\label{sm-fields}
\end{eqnarray}
where each superfield is assigned to the fundamental {\bf 4}
representation ($\alpha=1,\dots 4$ is the $SU(4)_H$ index). With this
field content the horizontal $SU(4)_H$ is anomalous. In order to
cancel the horizontal anomaly we introduce the following superfields
which are vectorlike with respect to $SU(3)\times SU(2)\times U(1)$
and belong to the ${\bf \bar 4}$ of $SU(4)_H$:
\begin{eqnarray}
F^\alpha: ~~ & &
U^\alpha\! \sim\! (3,1,2/3,\bar{4}),  ~~~
D^\alpha\! \sim\! (3,1,-1/3,\bar{4}),  ~~~
E^\alpha\! \sim\! (1,1,-1,\bar{4})      \cr
F_c^{\alpha}: ~~ & &
U_c^{\alpha}\! \sim\! (\bar{3},1,-2/3,\bar{4}),  ~~
D_c^{\alpha}\! \sim\! (\bar{3},1,1/3,\bar{4}),  ~~~
E_c^{\alpha}\! \sim\! (1,1,1,\bar{4}),   ~~
N_c^\alpha\! \sim\! (1,1,0,\bar{4})
%\hskip 3.5truecm \alpha=1,2,3,4
\label{new-fields}
\end{eqnarray}
%With respect to the vertical group $SU(3)\times SU(2)\times U(1)$
%the $F_c$'s transform in the same way as the $f^c$'s in (1), while
%the $F$'s have opposite charges with respect to the $F_c$'s.
As we will see in short, this same set of  superfields turn out to be necessary
also for providing masses to the known fermions.

In the Higgs sector, we choose the standard Higgs doublet superfields
$\Phi_{1,2}$ to be singlets under $SU(4)_H$. The additional Higgs
scalars needed for the breaking of the horizontal symmetry at some
large scale cannot couple to the standard $SU(2)\times U(1)$ gauge
bosons, and thus
must be singlets under the electroweak group.
In order to break completely the horizontal symmetry and to generate
realistic mass matrices for the fermions, we introduce a set of
$SU(3)\times SU(2)\times U(1)$ singlet `horizontal' superfields,
transforming either as the symmetric {\bf 10}
($\xi_{\{\alpha\beta\}}$) or as the antisymmetric {\bf 6}
($\chi_{[\alpha\beta]}$) representations of $SU(4)_H$.
Let us first consider the case with the fields $\xi^k$, $k=1,2\dots$
in symmetric representation. Additional superfields $\bar{\xi}^k$
% $\bar{\xi}^{\{\alpha,\beta\}}$ and $\bar{\xi}^{[\alpha,\beta]}$,
transforming as ${\bf \overline{10}}$ are   needed to render the
Higgsino sector free from chiral anomalies.
%while the {\bf  6} is (real and) automatically anomaly free.
However, these additional scalars do not contribute to the fermion
masses.\footnote{ Let us note that $\bar\xi^{\{\alpha\beta\}}$
in the ${\bf \overline {10}}$ cannot couple in renormalizable way
to the heavy vectorlike `matter' fields in the ${\bf \overline {4}}$,
and being an $SU(2)$ singlet, neither it can couple to quarks and
leptons. However, it is still possible to introduce a direct
non-renormalizable terms cutoff by the Planck scale
$(1/M_{Pl}) f_{\alpha} f^c_{\beta} \Phi_{1,2} \bar\xi^{\alpha\beta}$.
In this case, in order to reproduce the observed values of the fermion
masses, the $SU(4)_H$ symmetry should be broken at a scale very close
to $M_{Pl}$. On the other hand, as we will see in Sect. 4, the
phenomenology of the model requires a horizontal symmetry breaking
scale substantially smaller than $M_{Pl}$.
As a consequence, these non-renormalizable terms would provide only
negligible contribution to the fermion masses. }

What remains now to show, is that within the framework of the gauge
horizontal symmetry $SU(4)_H$ one can obtain a realistic mass pattern
for three families, ensuring at the same time that all fermions
of the fourth family are naturally heavy, say in the 100 GeV range.
Although we have started our considerations by a general analysis
which included also non-renormalizable operators as in
(\ref{mass-op}) and (\ref{Rvioloper}), in building the model we will
restrict ourselves to consider only renormalizable interactions.

The most general Yukawa superpotential for the down (up) quark and
for the lepton
superfields allowed by gauge invariance reads
\begin{equation}
W_F=g_f f_\alpha F_c^{\alpha} \, \Phi_{1(2)}
+ \sum_{k} h^k_F F^\alpha F_c^{\beta} \, \xi^k_{\alpha\beta}
+ \mu_f F^\alpha f^c_\alpha
%\hskip 1.truecm \alpha=1,2,3,4.
\label{W-fermions}
\end{equation}
with $f$ and $F$ respectively from (\ref{sm-fields}) and
(\ref{new-fields}). The analogous couplings for the neutrinos
have the form
\begin{equation}
W_N=g_\nu l_\alpha N_c^{\alpha} \, \Phi_{2}
+ \sum_{k} h^k_N N_c^\alpha N_c^{\beta} \, \xi^k_{\alpha\beta}.
%\hskip 1.truecm \alpha=1,2,3,4.
\label{W-neutrinos}
\end{equation}
\noindent
Here the $g$'s and $h$'s are Yukawa couplings which we assume to be
$O(1)$. The last term in eq. (\ref{W-fermions})
is a gauge invariant bilinear, and the $\mu_f$'s are gauge invariant large
mass parameters. As already stated, no terms trilinear in the
quark and lepton superfields are allowed by the
$SU(4)_H$ gauge symmetry, ensuring naturally the absence of the
B and L violating couplings $l q d^c$, $l l e^c$ and $u^c d^c d^c$.
We are facing here a situation analogous to the $SO(10)$ model,
since R-parity does not have to be imposed by hand,
but appears as an accidental symmetry
that follows from the requirement of horizontal gauge invariance.
Indeed, the superpotential is invariant with respect to a $Z_2$
transformation under which the fermion superfields $f,f^c,F$ and
$F_c$ (which have an odd number of $SU(4)_H$ indices) change sign,
while the Higgs superfields $\Phi_{1,2}$ and $\xi$ (with an even number of
$SU(4)_H$ indices) stay invariant.
More in general, the superpotential $W_F + W_N$
% of eqs. (\ref{W-fermions}) and (\ref{W-neutrinos})
has an automatic global symmetry $U(1)_H$
under the following transformations:
\begin{equation}
f,f^c\to e^{i\omega}f,f^c,~~~
F,F^c\to e^{-i\omega}F,F^c,~~~
\xi^k \to e^{2i\omega}\xi^k,~~~
\Phi_{1,2}\to \Phi_{1,2}  ~~~
(\bar{\xi}^k \to e^{-2i\omega}\bar{\xi}^k)
\label{U1H}
\end{equation}
where a $Z_2$ subgroup ($\omega\!=\!\pi$) remains unbroken even when
the scalars $\xi$ get non-zero VEVs. This $Z_2$ matter
parity ensures R parity conservation and hence proton stability.

The Yukawa couplings (\ref{W-fermions}) lead to the so called
"universal seesaw" mechanism \cite{uniseesaw} for the fermion mass
generation, which for the case of neutrinos reduces to the ordinary
seesaw mechanism \cite{seesaw}. Indeed, after the horizontal scalars
$\xi_k$ develop non-zero VEVs, the extra fermions $F$ and $F_c$ of
eq. (\ref{new-fields}) acquire large masses through the second term
in eq. (\ref{W-fermions}). Then the first and third terms cause a
``seesaw'' mixing of the ordinary quarks and leptons $f,f^c$ with
the heavy ones.
As a result, in the base $(f,F)$ $(f^c,F_c)$, the $8\times8$ mass
matrix for the down (up) type charged fermions $f=e,d,(u)$
reads \footnote{The scheme considered here is
a direct $SU(4)_H$ extension of the model
\cite{uniseesaw1} based on the horizontal symmetry $SU(3)_H$.
In the case when the vertical gauge symmetry is extended to the left-right
symmetric model $SU(2)_L\times SU(2)_R\times U(1)_{B-L}$, the values
of $\mu$'s are given by the scale of the $SU(2)_R$
breaking \cite{uniseesaw1}. }
\begin{equation}
{\cal M}_f=
\pmatrix{0              & g_f v_{1(2)} \cr
         \mu_f   &  \hat{M}^F \cr} \,,  ~~~~~~~
\hat{M}^F_{\alpha\beta} =
\sum_{k} h^k_F \, \langle\xi^k_{\alpha\beta}\rangle
\label{charged-mass}
\end{equation}
where $v_{1,2}=\langle \tilde \Phi_{1,2}\rangle$
are the VEVs of the two electroweak Higgs doublets.
%Since the off-diagonal blocks are generation blind (unit) matrices,
%the heavy fermion mass pattern is given essentially by the VEV
%structure of horizontal scalars:
As for the neutrinos, in the base $(\nu,N_c)$
%$[(\nu_e,\nu_\mu,\nu_\tau,\nu^\prime),\,(N_e,N_\mu,N_\tau,N^\prime)]$
the $8\times 8$ Majorana mass matrix has the form
\begin{equation}
{\cal M}_{\nu}=
\pmatrix{0              & g_\nu v_2 \cr
         g_\nu v_2  &  \hat{M}^N \cr} \,, ~~~~~~~
\hat{M}^N_{\alpha\beta} =
\sum_{k} h^k_N \, \langle\xi^k_{\alpha\beta}\rangle.
\label{neutral-mass}
\end{equation}
The universal seesaw picture provides a natural possibility to obtain
three light families, while the fourth one is heavy, say with
masses of the order of the electroweak scale.
Indeed, let us assume that the $4\times 4$ mass matrices
$\hat{M}_{F(N)}$ for the heavy fermions are rank-3 matrices of the
following form
\begin{equation}
\hat{M}_{F}=
\pmatrix{ M^{(3)}_F     &  0 \cr
               0  &  0    \cr}  ~, ~~~~~F=U,D,E,N
\label{degenerated}
\end{equation}
where the $3\times 3$ blocks $M^{(3)}_F$ contain non-zero entries.
In other words, we assume that all the VEVs of the type
$\langle \xi^k_{\alpha 4} \rangle $ are vanishing, so that
a diagonal $\tilde{U}(1)$ subgroup of $SU(4)_H\times U(1)_H$, given
by the generator $\tilde{T}=\mbox{diag}(0,0,0,1)$, is left unbroken.
In this case there is no seesaw mechanism for the fermions
of the fourth family: the right-handed components of the fields
$f_4=b',t',\tau',\nu'$ are actually the $F_c^4$ states, whereas the
$f^c_4$ form with the $F_4$ superheavy particles of mass $\mu_f$.

{}From eqs. (\ref{charged-mass}) and (\ref{neutral-mass}) we obtain
for the fourth family fermions
\begin{eqnarray}
& & m_{b'}=g_d v\cos\beta
% ~(>120\, \mbox{GeV}),
{}~~~~~~~~ m_{t'}=g_u v\sin\beta              \cr
& &  m_{\tau'}=g_e v\cos\beta        ~~~~~~~~
m_{\nu'}=g_\nu v\sin\beta
\label{fourth-mass}
\end{eqnarray}
where $v=174\,$GeV is the electroweak breaking scale and
$\tan\beta= v_2/v_1$.
% and we have given inside the brackets
% the experimental limits on the fourth generation masses.
Since all the Yukawa
couplings are assumed to be $O(1)$, for moderate values of
$\tan\beta$ all the masses in (\ref{fourth-mass}) are
of the order $\sim$ 100 GeV.
On the experimental side,
the firmest constraints on the masses of any new sequential fermion,
quark or lepton, have been set at LEP: $ m_f \gtrsim M_Z/2$.
This indeed represent the best constraint on
$m_{\tau^\prime}$ and $m_{\nu^\prime}$.
Searches for new quarks at the TEVATRON collider could in principle
give much better bounds for
$m_{t^\prime}$ and $m_{b^\prime}$\cite{bprime-limits}.
However, let us note that the structure of the heavy mass matrix
(\ref{degenerated}) implies that the fourth family is unmixed with the
three lighter ones. Hence the usual signatures, as
for example $b^\prime \to c,u$,
that have been used to set the limits on new sequential quarks
\cite{bprime-limits} do not occur in our case. In the absence of a
detailed experimental analysis of the unmixed case, the only reliable
limit is again the LEP one also for the new quarks. Hence we can safely
conclude that the predictions in (\ref{fourth-mass}) are by no means in
conflict with the existing experimental limits.
However, it is clear that for
the masses of the fourth family fermions not much room is left.
The allowed parameter space is in fact strongly constrained by
the CDF measurement of the top mass,
$m_t=174\pm 10\pm 13\,$GeV \cite{CDF},
by the precision tests of the SM which do not
leave much space for additional sizeable radiative
corrections as would be induced by a too large
$m_{t^\prime}$-$m_{b^\prime}$ splitting, and by
renormalization group
(RG) analysis of the Yukawa couplings, much in the spirit
of ref. \cite{fourth-susy}.

In particular, while the general analysis in \cite{fourth-susy}
does allow for the possibilities $m_{t^\prime} < m_t$
or $m_{t^\prime} < m_{b^\prime}$, in our model these pattern of masses
are not allowed. In fact the universal seesaw mechanism
outlined before implies $m_{t^\prime} \geq  m_t$, and most likely
$m_{t^\prime} \geq m_{b^\prime}$.
Then, according to \cite{fourth-susy}, for
$m_{t^\prime} \geq  m_t > 150\,$GeV the consistency of
the model implies not too large values for the
masses of the other fermions in fourth family.
Namely,  for the low values of $\tan\beta$ we are interested in
(e.g. $\tan\beta \sim 2$), the maximal values allowed are about
$m_{b^\prime}\sim 100\,$GeV and
$m_{\tau^\prime,\nu^\prime} \sim 50\,$GeV,
that is within the reach of LEP II.

Let us now consider the mass matrices for the first three families.
In this case the seesaw mechanism is  effective for
suppressing the fermion masses from the electroweak scale down
to the observed values.
By assuming $M_F^{(3)}> \mu_f$, it is apparent from
(\ref{charged-mass}) that the fermions of the first three families
will acquire their masses through a mixing with the superheavy $F$
fermions. Namely, after decoupling the heavy states, the $3\times 3$
mass matrices of the light charged down (up) type fermions are
\footnote{ After decoupling the heavy states at the
horizontal symmetry scale $V_H$, our model simply reduces
to the SSM with four families. In fact, eqs. (\ref{fourth-mass}),
(\ref{mass-light}) and (\ref{mass-neutrino}) define the fermion
running masses at $\mu=V_H$. In order to deduce the fermion physical
masses the RG running has to be taken into account. }
\begin{equation}
m_f^{(3)} = g_f \mu_f (M_F^{(3)})^{-1} v_{1(2)},~~~~~~~f=d,e,(u)
\label{mass-light}
\end{equation}
while the $3\times 3$ Majorana mass matrix for the light neutrinos
obtained from (\ref{neutral-mass}) reads
\begin{equation}
m_\nu^{(3)} = (M_N^{(3)})^{-1} (g_\nu v_2)^2.
\label{mass-neutrino}
\end{equation}

In contrast to the SM and to most GUT models, in our picture the
fermion mass hierarchy is not generated by an {\em ad hoc} choice
of the Yukawa coupling constants.
In fact, in our scheme all the Yukawas are assumed to be of the same
order of magnitude, for example $O(1)$ or close
to the size of the gauge couplings.
As long as the off-diagonal blocks in eqs. (\ref{charged-mass}) and
(\ref{neutral-mass}) are flavour blind (unit) matrices, all the
informations on the fermion mass and mixing pattern is contained
in the heavy fermion mass matrices $M^{(3)}_F$.
%In order to obtain the observed mass hierarchy among the fermion
%families together with naturally small mixings, the heavy fermion
%mass matrices $M^{(3)}_F$ should have a hierarchical pattern.
Since the structure of the latter is determined by the different VEVs
$\langle \xi^k \rangle$ (modulo differences in the Yukawa constants
$h_F^k$), the observed hierarchy of the light fermion masses is
ultimately determined by the hierarchy in the VEVs which break the
horizontal symmetry. In other words, the VEV pattern should provide
the step-by-step breaking of the chiral horizontal symmetry
\begin{equation}
SU(4)_H\times U(1)_H\stackrel{V_1}{\rightarrow}
SU(3)_H\times U(1)'_H\stackrel{V_2}{\rightarrow}
SU(2)_H\times U(1)''_H\stackrel{V_3}{\rightarrow} \tilde{U}(1)
\label{chain}
\end{equation}
so that the first breaking (at the scale $V_1\sim\langle\xi_{11}\rangle$)
defines the mass terms for the first heavy family $F_1$, the second
breaking (at $V_2\sim \langle\xi_{12}\rangle, \langle\xi_{22}\rangle$)
for the second family $F_2$ etc. Through the seesaw mechanism this
horizontal VEV hierarchy is reflected in the observed pattern of fermion
masses. Namely, from (\ref{mass-light}) and (\ref{mass-neutrino})
it is clear that the hierarchy among the light families is
inversely proportional to the one between the heavies \cite{uniseesaw1}
(see also \cite{Rattazzi}),
while the unbroken global symmetry $\tilde{U}(1)$ ensures the natural
heaviness of the fourth family fermions.

Let us now briefly analyse the issue of horizontal symmetry breaking.
The simplest way to achieve non-zero VEVs for the scalars $\xi$
is to introduce a set of superfields $S_a$ and $\Sigma_a$, respectively
in $SU(4)_H$ singlet and adjoint representations
($a=1,2\dots$), and to consider, according to our `renormalizability'
paradigm, the general superpotential
%\begin{equation}
%W_H= M_k\xi_k \overline{\xi}_k +
%\lambda_{akl} S_a \xi_k \overline{\xi}_l + P(S_a)
%\label{WH}
%\end{equation}
\begin{equation}
W_H= M_k\xi_k \overline{\xi}_k +
\lambda_{akl} S_a \xi_k \overline{\xi}_l +
\lambda'_{akl} \xi_k \Sigma_a \overline{\xi}_l + P(S,\Sigma)
\label{WH}
\end{equation}
where $P(S,\Sigma)$ is a general $3^{rd}$ order polynomial of the $S_a$
and $\Sigma_a$ fields (containing linear, bilinear and trilinear terms).
Notice that this superpotential automatically respects the $U(1)_H$
invariance (\ref{U1H}), but has no additional accidental global
symmetries.

The superpotential (\ref{WH}) in itself does not break SUSY.
% both $F$ and $D$ terms are vanishing.
Moreover, in the exact SUSY
limit the vacuum state is highly degenerated -- there are
several zero-energy vacua with different configurations of horizontal VEVs.
It would be a difficult task to provide an exhaustive
analysis of all the possible
vacua in the general case, that is to decide which configuration
of VEVs is chosen as the true vacuum once the soft SUSY breaking terms
are included. However, taking into account that after SUSY
breaking the potential of the horizontal scalars has to a large extent
the general structure of usual (non-SUSY) Higgs polynomial,
one can argue that for a certain choice of parameters it is possible
to obtain
the needed pattern of VEVs (see for example the analysis in
refs. \cite{Khlopov,Mamuka} for the case of $SU(3)_H$ symmetry).

Indeed, let us consider a first case with {\em only} one pair of
$\xi_1+\bar{\xi}_1$ superfields, the ones which have the largest VEV
$(V_1)$ in the exact SUSY limit. The constraint from the $D-$term tells us
that in this case $\langle\xi_1\rangle = \langle \bar{\xi_1} \rangle$.
Then it is easy to show that after SUSY breaking, for a proper
choice of the range of values for the relevant parameters,
the true vacuum can have the
configuration $\langle\xi_1\rangle = V_1\mbox{diag}(1,0,0,0)$
which breaks $SU(4)_H\times U(1)_H$ down to $SU(3)_H\times U(1)'_H$.
Therefore, at this stage only the first family of $F$ fermions
gets a mass through the couplings (\ref{W-fermions}) while
the others, being protected by the residual chiral symmetry
$SU(3)_H\times U(1)'_H$, remain massless.\footnote{
Alternatively, for the complementary choice of the
parameter range, one would have the vacuum
$\langle \xi_1 \rangle \propto \mbox{diag}(1,1,1,1)$ which breaks
$SU(4)_H\times U(1)_H$ down to $SO(4)_H$. This pattern, however,
does not maintain chirality and leads to degenerate fermion masses.}

In analysing the scalar potential of the fields $\xi_2+\bar{\xi}_2$ with
next largest VEV ($V_2$), we have to take into the account that after
`decoupling' $\xi_1$ (i.e. substituting $\xi_1\to\langle\xi_1\rangle$),
the symmetry group is reduced to $SU(3)_H\times U(1)'_H$, under which
$\xi_2$ branches as {\bf 10}={\bf 6}+{\bf 3}+{\bf 1}.
The VEVs which give masses to the second heavy generation
belong to the {\bf 6} and {\bf 3}, and can be chosen in the form
$\langle\xi_{22}\rangle$ and $\langle\xi_{12}\rangle$ respectively.
These break $SU(3)_H\times U(1)'_H$ down to $SU(2)_H\times U(1)''_H$,
thus respecting a residual  chiral symmetry for the third and fourth
family of heavy fermions which at this stage remain massless.
Finally, yet another pair $\xi_3+\bar{\xi}_3$ can develop VEVs
in the $\alpha 3$ directions ($\alpha=1,2,3$) thus breaking
$SU(2)_H\times U(1)''_H$ down to $\tilde{U}(1)$, which acts only
on the fourth family.

One could try to avoid introducing
the adjoint representations $\Sigma_a$ and
keep in the superpotential (\ref{WH}) only the singlet fields $S_a$.
Then, in the exact SUSY limit the vacuum state would have a
continuous degeneration -- there will be vacuum valleys.
%-- superpotential would fix the absolute values of the horizontal VEVs,
%but not their orientation.
The reason for this is that in this case
the superpotential would have an
accidental global symmetry $SU(10)$ larger than local $SU(4)_H$.
In the SUSY limit these valleys correspond to massless Goldstone
modes given by certain components of the horizontal superfields.
When SUSY breaking is taken into account, the radiative corrections
explicitly break the extra global symmetry, lifting the
vacuum degeneracy and providing $\sim 100$ GeV masses to the horizontal
Goldstone modes, which would then become pseudo-Goldstone, massive
familon-like
scalars. In general these states would
have diagonal as well as non-diagonal
Yukawa couplings with the fermions, and
in particular the strength of the couplings to
the light fermions would be suppressed by a factor $\sim v/V_H$.
%where $V_H$ stands for the horizontal symmetry breaking scale.

It is a very difficult task to provide a full analysis of the VEV pattern in
this case and deduce which configuration of VEVs is fixed as the
true vacuum state after SUSY breaking.
Namely, already for two pairs of $\xi+\bar{\xi}$ the general
VEV structure of vacuum valleys {\em completely} breaks the $SU(4)_H$
symmetry, not maintaining the $\tilde{U}(1)$ subgroup.
%of the $SU(4)_H\times U(1)_H$.
%(which we needed to ensure the structure
%(\ref{degenerated}) providing naturally heavy fourth family).
One can still argue that for a proper choice of the relevant parameters
the needed pattern of VEVs can be obtained. Though this possibility
can be interesting from the phenomenological point of view, it
deserves a special investigation. Therefore, in the following
we will assume that there are no light familon-like scalars and that all
the horizontal fields have $O(V_H)$ masses.

As we have mentioned at the beginning of this section,
in order to generate masses for the heavy states $F$ and $F_c$
it is also possible to introduce horizontal
Higgs fields $\chi_{[\alpha\beta]}$ in the
antisymmetric {\bf 6} representations of $SU(4)_H$.
The $\langle \chi_k \rangle$ VEVs
would then contribute to the mass matrices of the heavy
{\it charged} fermions through the term
$\sum_{k} h^k_F F^\alpha F_c^{\beta} \, \chi^k_{[\alpha\beta]} $,
while the corresponding term for the
Majorana mass matrix of the heavy neutral states
$N^c$ is forbidden due to the antisymmetry of the representation.
However, in this case the appearance of the terms like
$M^\prime_k \chi_k \chi_k$
in the superpotential for the horizontal fields would break
explicitly the global $U(1)_H$ in eq. (\ref{U1H})
and hence the
residual $\tilde U(1)$ invariance, thus rendering
unnatural the degenerate structure of the heavy matrices
$\hat{M}_{F}$ in eq. (\ref{degenerated}). Although
for $\langle \chi_k \rangle \leq \mu_f$
the heaviness of the fourth
family charged fermions would still be guaranteed,
we would lose
a natural explanation for a heavy $\nu^\prime$.
In fact, through additional
terms like $\xi\, \Sigma\, \chi$
sizable VEVs in the $\,\alpha 4$ directions
would be induced also for  the $\xi$ fields.
If, as it seems natural to occur, the induced VEVs are larger than
the electroweak scale,  the fourth family neutrino mass
will also result from a seesaw giving
$m_{\nu^\prime} \sim (g_\nu v_2)^2/ \langle \xi \rangle_{\rm induced}$.
That is, the $\nu^\prime$
will also be light, thus rendering the model phenomenologically
unacceptable.

%the terms like $\xi_{\alpha\beta}\bar{\xi}^{\alpha\gamma}
%\xi_{\gamma\delta}\bar{\xi}^{\delta\beta}$ arise (though with small
%constants), which absent in SUSY limit. in the case of non-SUSY
%general potential there exists a parameter space when the needed
%pattern of VEVs can be obtained

%The following remark on the VEV hierarchy is in order.
%The true vacuum will be defined only after SUSY
%breaking. By a proper choice of the parameter range we can always achieve
%that the true vacuum is given by the first configuration
%\begin{equation}
%\langle \xi \rangle = V_{11} \mbox{diag}(1,0,0,0)\,,~~~~~
%V_{11}\sim \sqrt{\mu_1 M}
%.\label{vac}
%\end{equation}

%This superpotential also obeys the global $U(1)_H$ symmetry,
%but has no more extra global symmetry exceeding the local one.

%%%%%%%%%%%%%%%%%%%%%%%%%%%%%%%%%%%%%%%%%%%%%%%%

\section{Phenomenological consequences of the model }

As was discussed above, the quark and charged lepton masses
at the scale $V_H$ are given by eqs. (\ref{fourth-mass}) and
(\ref{mass-light}). By assumption, the heavy fermion mass matrices
$M^{(3)}_F$ are non-degenerate, and thereby have three massive
eigenstates, with mass hierarchy reflecting the $SU(4)_H$ symmetry
breaking pattern. The weak mixing angles are determined by the
structure of these matrices, whereas the quark and lepton masses
are inversely proportional to the masses of their heavy partners.
For the down quark and charged lepton masses we have
\begin{equation}
m_{d,s,(b)}\simeq \eta_{d,(b)} m_{b^\prime}
\frac{\mu_d}{M_{D,S,(B)}} < m_{b^\prime}\,,~~~~~
m_{e,\mu,(\tau)}\simeq \eta_{e,(\tau)} m_{\tau^\prime}
\frac{\mu_e}{M_{E,\cal{M},(\cal{T})}} < m_{\tau^\prime}
\label{mbtau}
\end{equation}
where $D,S,B$ ($E,\cal{M},\cal{T}$) are the mass eigenstates of
$M^{(3)}_{D(E)}$, and the factors $\eta$ account for the
differences in the RG
running of masses from the horizontal scale to lower energies.
The fact that the $b$ and $\tau$ masses are of order
a few GeV, implies that
the masses of the corresponding heavy states $B$ and $\cal{T}$
are not much larger (say, within one or two orders of magnitude)
than the mass scale $\mu_{d,e}$. As for the top quark, the value of
its mass $m_t\gtrsim 150\,$GeV requires $M_T\sim \mu_u$.
In this case corrections to the seesaw formula (\ref{mass-light})
should be taken into account in relating $m_t$ to the heavy scales
(see e.g. ref. \cite{Rattazzi}).

As a result of the seesaw mechanism for the fermion masses generation,
the light charged states correspond to some superposition of the
($f$,$f^c$) and ($F$,$F^c$) states. It is well known that a mixing
between the light $SU(2)$ doublet states $f$ and the heavy $SU(2)$
singlets $F$ could induce FCNC in the electroweak interactions and will
also alter the flavor diagonal couplings of the light states \cite{fit}.
However, in our case such a mixing is suppressed as the ratio
$v_{1,(2)}/\hat{M}_F$ and thus negligibly small when compared
with the present experimental bounds \cite{fc-limits,fitnew}.
On the other hand, the mixing between the  $SU(2)$ singlets $f^c$ and
$F^c$ can be large, since it is controlled by the ratio of the two
mass scales $\mu_f$ and $\hat{M}_F$ which, as we have seen, can be as
large as $\sim 10^{-1}$  or even close to unity in the case of the
$t$ quark. However, this kind of mixing between states transforming
in the same way under $SU(2)\times U(1)$ cannot affect the electroweak
quantities, and is essentially unobservable.

Let us discuss now the neutrino masses. As we have already stated,
three neutrinos $\nu_{e},\nu_\mu,\nu_\tau$ are light Majorana
particles. Their running masses at $\mu=V_H$ are determined by
the heavy 'right-handed' neutrino eigenstates $N_e,N_\mu,N_\tau$ at
their decoupling, according to the seesaw formula (\ref{mass-neutrino}).
As for the fourth neutrino $\nu'$, it appears to be
a heavy Dirac particle with mass $\sim 100\,$GeV.
Then for the neutrino physical masses we have
\begin{equation}
m_{\nu_e,\nu_\mu,\nu_\tau} = \eta_\nu\,\frac{(m_{\nu'})^2}
{ M_{N_e,N_\mu,N_\tau} }\,,~~~~~~~
m_{\nu^\prime} \geq \frac{M_Z}{2}
%= R_{\nu^\prime} \bar{m}_{\nu'}
\label{seesaw}
\end{equation}
where the factor $\eta_\nu$ accounts for the different RG running
of Majorana and Dirac masses from the $SU(4)_H$ breaking scale to
lower energies (for the RG running of Majorana neutrino masses see
e.g. \cite{BLP}). Therefore,
modulo the different Yukawa couplings $h_F$, the neutrino mass
hierarchy is expected to be qualitatively the same as the hierarchy
between the quarks or the charged leptons:
\begin{equation}
m_{\nu_e}: m_{\nu_\mu}:m_{\nu_\tau} \sim m_u : m_c :m_t ~~~~~
{\rm or} ~~~~~m_e:m_\mu : m_\tau
\label{neutrino-hierarchy}
\end{equation}
%For the heaviest among those, namely the $\tau$-neutrino, we have
%$m_{\nu_\tau}=....$.
Below the scale $V_H$ our theory is just the SSM, and all FCNC
phenomena related with the horizontal symmetry are strongly suppressed.
Therefore, all neutrinos are effectively stable on a cosmological
scale. In order to respect the cosmological upper bound \cite{GZ} on
the light stable neutrino masses
%implied by the measured age of the Universe
$m_{\nu_\tau} \leq 92\,\Omega\, h^2$ eV
(here $\Omega=\rho/\rho_c \sim 1$ is the ratio of the energy density
of the Universe to the critical density, and $h=0.4-1$ is the Hubble
parameter in units of $H_0=100\,$Km s$^{-1}\,$Mpc$^{-1}$)
the following lower bound on the mass of $N_\tau$
must be respected :
%smallest of the
%three nonzero eigenvalues of $\hat{M}^N$, namely $M_{N_\tau}$:
%\footnote{We note that FCNC couplings induced by the non-diagonal
%elements in $M_N$ (\ref{seesaw}) would in fact allow, via $Z_0$ exchange,
%for the decay $\nu_\tau \to \nu_\alpha \nu_\beta \overline{\nu_\beta}$
%with $\alpha, \beta = e$ or $\mu$. However, being these couplings
%suppressed as $m_{\nu^\prime}/M_N \lesssim 10^{-10}$ all the three
%light neutrinos are expected to be effectively stable on a
%cosmological scale.}
\begin{equation}
M_{N_\tau} = h_N V_H \geq \frac{\eta_\nu}{4\Omega h^2}
%\left(\frac{R}{R'}\right)^2
\left(\frac{m_Z^2}{92\, \mbox{eV}}\right)\gtrsim 10^{11}~ \mbox{GeV}.
\label{univ_age}
\end{equation}
In deriving this limit we have taken into account that the present
age of Universe $t_0\gtrsim 12\cdot 10^{9}$ yr ~ requires, for
$\Omega=1$, the Hubble parameter $h\simeq 0.5$, and we have assumed that
$\eta_\nu \sim 1$. As long as the Yukawa constants $h_N$ are $O(1)$,
this bound translates into a lower bound on the smallest scale
$V_H=V_3$ in the $SU(4)_H$ symmetry breaking chain (\ref{chain}).

Let us now address some phenomenological issues regarding the fourth
family fermions. The structure of the heavy mass matrix
(\ref{degenerated}) which leaves unbroken the diagonal
$\tilde{U}(1)$ subgroup of $SU(4)_H\times U(1)_H$, also ensures that
the fourth family is unmixed with the three lighter ones.
We assume that the lightest member of the fourth generation is
the neutral one $\nu^\prime$, as is also suggested by the analysis
of ref. \cite{fourth-susy}. For simplicity we also assume that
$m_{t^\prime} > m_{b^\prime}$.  Then  $b^\prime$ and $\nu^\prime$
are stable with respect to electroweak interactions.

The presence of stable neutrinos $\nu^\prime$ with mass in the
100 GeV range is phenomenologically and
in particular cosmologically acceptable,
since their contribution to the cosmological energy density is
vanishingly small.
Only in the presence of a sizeable
$\nu^\prime$-$\overline{\nu^\prime}$ primordial asymmetry
the stable relics $\nu^\prime$ would contribute to the present cosmological
density,  and this contribution would still be
acceptable as long as
their present number density $n_{\nu^\prime}$ does not exceed
the baryon number density $n_B$.
However, as we will argue in the following,
in our model no sizeable asymmetry has to be expected
for the fourth family fermions.

In contrast, the existence of stable heavy quarks carrying colour
and electric charge would constitute a potential problem for the model,
since it will conflict with the constraints arising from superheavy
element searches, as well as with other cosmological and astrophysical
constraints \cite{heavy,champ}. Indeed, the stable $b^\prime$ would
behave essentially as $d$ quarks, hadronising into heavy `protons' and
giving rise to heavy hydrogen-like `isotopes' with masses
$\sim 100$ GeV. The existing experimental limits on this kind of
isotopes are extremely tight. For example for masses $m_{b^\prime}< 1$ TeV
the limit on their abundance relative to normal hydrogen
is $n_{b^\prime}/n_B < 10^{-28}$  \cite{heavy-H}.

However, the exchange of the $SU(4)_H$ gauge bosons $Z_H$ would
allow the heavy quark to decay, dominantly  through the channel
$b^\prime \to b \bar{\nu}_{\tau} \nu^\prime $, with a lifetime
\footnote{Much faster decay $b^\prime \to b+J$ can occur if there
are pseudo-Goldstone familon-like scalars $J$, arising from the
breaking of accidental global $SU(10)$ symmetry of the superpotential
(see the discussion in Sect. 3).
However, here we will not consider such a possibility. }
\begin{equation}
\tau_{b^\prime}\sim \left(\frac{V_H}{v}\right)^4
\left(\frac{m_\mu}{m_{b^\prime}}\right)^5
\tau_\mu =
\left(\frac{V_H}{10^{12}~ \mbox{GeV}}\right)^4
\left(\frac{150~\mbox{GeV}}{m_{b^\prime}}\right)^5
\cdot 4\cdot 10^{17}~\mbox{s}
\label{lifetime}
\end{equation}
where $V_H=V_3$ is the lowest scale in the horizontal symmetry
breaking (see eq. (\ref{chain})), $v$ is the electroweak scale
and $\tau_\mu = \tau(\mu\rightarrow e\bar{\nu}_e \nu_\mu)
\simeq 2.2\times 10^{-6}\,$s. is the muon lifetime.
We can use cosmological arguments, together with the
experimental limits on searches of heavy isotopes, to put an upper
bound on $\tau_{b^\prime}$, which in turn will translate in an upper
limit on $V_H$. Indeed,
taking into account the finite lifetime of the heavy quarks,
their present number abundance relative to baryons is
$\,n_{b^\prime}/n_B = r_0 \exp({-t_0/\tau_{b^\prime}}) \leq 10^{-28}$,
%\begin{equation}
%\frac{n_{b^\prime}}{n_B} = r_{b^\prime}
%\frac{m_B}{m_{b^\prime}}\frac{\Omega_{b^\prime}}{\Omega_B}
%\exp({-t_0/\tau_{b^\prime}})
% \leq 10^{-28},
%\label{tau}
%\end{equation}
where $r_0=(n_{b^\prime}/n_B)_0$ represents the relic
abundance for stable $b^\prime$.
%and $t_0 \simeq 2\cdot 10^{17}h^{-1}\,$s is the
%present age of the Universe.
{}From this equation we get the upper limit on the $b^\prime$ lifetime
\begin{equation}
\tau_{b^\prime}\leq 3 \cdot 10^{15} h^{-1}
(1+ 0.036\,\lg r_0)^{-1}\, \mbox{s}.
\label{tau}
\end{equation}
One cannot say definitely what is the value of $r_0$, due
to many theoretical uncertainties related to the actual annihilation
cross section for the $b^\prime$.
However, an estimate of the relic abundance
of heavy stable $d$-type quarks has been given in \cite{heavy}.
Under the assumption that there is no cosmological baryon asymmetry
between the  $b^\prime$ and  $\bar{b}^\prime$,  it was found that for
$m_{b^\prime} \sim 150\,$GeV
the energy density of these relics, relative to critical density
(namely $\Omega_{b^\prime} h^2$ )
could range from $10^{-9}$  to
$10^{-4}$ (smaller values are obtained for lighter $b^\prime$
masses). The lower limit corresponds to the case when the relic
density is determined by the annihilation after the QCD phase
transition, and it was obtained by taking as an upper bound
on the annihilation cross section
the geometrical cross section
($\sigma_0\sim 100\,$mb ).
The upper limit was obtained under the opposite assumption, namely that
annihilation after confinement is negligible, and that
the relic density is essentially determined by the QCD annihilation
cross section for free quarks.
Then the ratio of the $b^\prime$ to baryon number densities
$r_0 = (n_{b^\prime}/n_B)_0 =
\Omega_{b^\prime} /\Omega_B \cdot m_B/m_{b^\prime}$
lies in the range $3\cdot 10^{-10} < r_0 < 3\cdot 10^{-5}$,
where we have taken $\Omega_B \sim 0.02$ as suggested
by nucleosynthesis estimates.
As is discussed in \cite{heavy}, the most
reasonable assumption is that the relevant annihilation process happens
after confinement, however with a cross section much smaller than the
geometrical one, giving $r_0 \sim 10^{-7}-10^{-8}$.
Clearly, in the presence of a sizeable baryon asymmetry in the fourth family
sector, the relic abundance of the heavy $b^\prime$ quarks would be
some orders of magnitude larger than the quoted estimates.

As we see, the bound (\ref{tau}) very weakly depends on the
initial $b^\prime$ abundance. Even if
we allow for a large primordial asymmetry
for the $b^\prime$, and let $r_0$ range between
$1-10^{-10}$, by taking $h=0.5$
we obtain
$\tau_{b^\prime}\leq 6\cdot 10^{15}-10^{16}\,$s.
%as required by the Universe age
%$t_0\geq 12\cdot 10^{9}\,$yr for $\Omega=1$.
On the other hand, according to eq. (\ref{lifetime}),
this bound translates into an extremely strong
upper limit
\begin{equation}
V_H \leq\, \frac{ (m_{b^\prime}/150~\mbox{GeV})^{5/4} }{ h^{1/4}
(1+ 0.036\,\lg r_0)^{1/4} }\cdot 3\cdot 10^{11} ~\mbox{GeV}
\leq  4 \cdot 10^{11} ~\mbox{GeV}
\label{V_H}
\end{equation}
 This bound corresponds to
$m_{b^\prime}<150\,$GeV, and the scale $V_H$ cannot much exceed  this limit
unless $m_{b^\prime}\gg 200\,$GeV.
For the more realistic values $m_{b^\prime}\simeq 100\,$GeV suggested by the
analysis in
\cite{fourth-susy} we get $V_H\leq 2.4\cdot 10^{11}\,$GeV.

More stringent limit on $r_0$ and $\tau_{b^\prime}$ can be derived
by considering that the late decay of the $b^\prime$ can cause a
significant contribution to observed cosmic ray fluxes, in particular
to the isotropic diffuse gamma-ray background \cite{Berez}.
Indeed, at the moment of decay,
the $b^\prime$ quarks are bounded within colorless hadrons like
$b^\prime u d$ or $\overline{b^\prime} u$ \cite{heavy}.
Then in the decay $b^\prime \to b \overline{\nu}_{\tau} \nu^\prime$
an unstable hadronic state emerges with the excitation energy
$E_0\simeq \frac{1}{3}m_{b^\prime}$. This will essentially appear
as a hadronic jet with the $b$ quark being the leading particle. The
fragmentation of this jet produces $\pi^0$, $\eta$ etc., with
the subsequent radiative decay resulting in a specific photon spectrum.
Obviously, the amount of produced photons is directly proportional
to $r_0$. In order to estimate their flux in the present Universe,
the redshift of their energies has to be taken into account as well.
As long as the decay happens at the matter dominated epoch, and the small
amount of relativistic decay products does not affect sensibly the
Universe expansion rate, we have
$1+z=(t_0/\tau_{b^\prime})^{2/3}\sim 10-20$
for the values of $\tau_{b^\prime}$ estimated above.
We also need to know what fraction of the
jet energy $E_0$ is taken by the photons and what is the energy
spectrum. These issues were studied in ref. \cite{Salati}, where
the photon spectra produced at jet hadronization were computed
for different leading particles using a Monte Carlo simulation
program \cite{Monte-Carlo}. It was shown that these spectra
%computed at different jet energies
exhibit a remarkable scaling property in terms of the
variable $x=E_\gamma/E_0$,
and in the case of leading particle being a $b$ quark, the photons
carry away about 25 percent of the initial jet energy.
Using the results of ref. \cite{Salati}
%for the photon spectrum $dN_\gamma/dx$ due to jet fragmentation,
we have computed the value of the isotropic cosmological gamma-flux
$d\Phi_\gamma/dE_\gamma$ and we have
compared it with the existing observational limits (see
\cite{Berezinsky} and references therein). For example,
for $E_\gamma=100$ MeV
the experimental upper bound on the $\gamma$-flux is of about
$10^{-7}\,$cm$^{-2}\,$s$^{-1}\,$sr$^{-1}\,$MeV$^{-1}$. We have
obtained that the cosmic gamma-flux produced due to $b^\prime$ decay
at $z=10-20$, saturates the above bound for $r_0\sim 10^{-7}$
which is close to, but still not in conflict
with our estimate of the $b^\prime$ relic abundance
in the $b^\prime$-$\overline{b^\prime}$ symmetric case.
Substantially larger $r_0$ would require much larger redshift, and
hence much smaller $\tau_{b^\prime}$. On the other hand,
%the cosmological limit on the $\tau$-neutrino mass
the lower bound (\ref{univ_age}) on the horizontal symmetry
breaking scale $V_H$ already excludes much smaller lifetimes.

This analysis implies that $r_0$ should be rather small, so
that any sizeable cosmological baryon asymmetry between
$b^\prime$ and  $\bar{b}^\prime$ is excluded.
This severely constrains the possible baryogenesis
mechanisms applicable to our model.
%that could produce the observed baryon asymmetry.
The appearance of baryon asymmetry in the fourth family in
itself is hardly expected, since it is unmixed with the other
three families and hence it has no source of CP violation.
However, the sphaleron effects \cite{KRS,Arnold} would immediately
redistribute the baryon asymmetry produced within the first three
families to the fourth family fermions. Therefore, no mechanism
is acceptable which generates the baryon asymmetry before the
sphaleron effects are switched off, that is
before the electroweak phase transition.\footnote{In principle,
in our model the baryogenesis with non-zero B$-$L could occur
%by Fukugita-Yanagida mechanism \cite{FY},
due to CP violation effects in out-of-equilibrium decays
$N^c \to l+ \Phi$ of the heavy right-handed neutrino \cite{FY} (for
the viability of this mechanism in the SUSY case see
ref. \cite{Olive}), or in the decays of $SU(4)_H$ gauge or scalar
bosons. Then sphaleron effects would immediately transfer the
produced net lepton number into a baryon asymmetry also in the
fourth family sector. Fortunately, our model naturally avoids the
possibility of such a lepto-baryogenesis.
As it was shown in ref. \cite{Olive}, the large scale density
fluctuations hinted by the COBE measurements require rather
low inflationary reheat temperature ($T_R\sim 10^8\,$GeV) and
correspondingly low inflaton mass ($m_\eta\sim 10^{11}\,$GeV).
On the other hand, the lower bound (\ref{univ_age}) on $M_N$ (and
respectively on $V_H$) tells us that masses of the right-handed
neutrinos and horizontal bosons should exceed $10^{11}\,$GeV,
and therefore they are not produced after inflation. }
In the context of our model the most appealing
possibility is to to assume that no baryon asymmetry is produced
before the electroweak epoch, and baryogenesis takes place at the
electroweak (first order) phase transition.
Such a baryogenesis mechanism is associated with the walls of the
expanding bubbles of the broken phase \cite{Misha}. Outside of
the bubbles electroweak symmetry is unbroken, quarks are massless
and the rate of the fermion number violation due to
sphaleron transitions greatly exceeds the Universe expansion rate.
%so that the sphaleron processes are in thermal equilibrium.
Inside the bubbles the quarks are massive due to non-zero VEVs of the
Higgs fields, while the sphaleron processes are strongly suppressed
and fermion number is effectively conserved. Then baryon asymmetry
inside the bubbles could be produced (and maintained) due
to CP violating effects, as a difference between the quark and
anti-quark fluxes penetrating the walls from the unbroken phase to the
broken one. Obviously, this concerns only the first three family
fermions. Since the fourth family is unmixed, has no CP violation,
and moreover all the fermions are very heavy, no baryon excess is
expected in this sector. Although the viability of such a
baryogenesis in the SM is still disputed
in the literature \cite{Gavela}, in the context of SSM it could be
more effective and sufficient for providing the observed baryon asymmetry.
Clearly this topic deserves additional special considerations.

According to eq. (\ref{seesaw}),
the upper limit on the horizontal symmetry breaking
scale $V_H \lesssim 4 \cdot 10^{11}\,$GeV
together with the experimental limit $m_{\nu^\prime} \geq M_Z/2$
translates into a
lower bound on the $\tau$-neutrino mass:
\begin{equation}
m_{\nu_\tau}\geq \,\frac{\eta_\nu}{h_N}\,
\frac{m_Z^2}{4V_H} \gtrsim (1-10)~\mbox{eV}
\label{nutau}
\end{equation}
where in the numerical estimate we have taken into account the $O(1)$
uncertainties in the relative renormalization factor $\eta_\nu$
and in the Yukawa coupling $h_N$
(for perturbativity we have to assume $h_N<3$ at $\mu=V_H$).
% which, as we have already
%mentioned, is compatible with a $\nu_\tau$ mass of the order
%of a few eV, while disfavour much lighter masses.
A $\nu_\tau$ with mass in the range $1-10$ eV will give a sizeable
contribution to the cosmological energy density as a hot dark matter
(HDM) component, while according to
(\ref{neutrino-hierarchy}) $\nu_\mu$ and $\nu_e$
are expected to have much smaller masses.
We remind here that the COBE measurements of the
cosmic microwave background anisotropy, together with other data on
the density distribution of the Universe at all distance scales
(galaxy-galaxy angular correlations, correlations of galactic
clusters, etc.) can all be fit by assuming some HDM admixture
to the dominant CDM component \cite{Shafi}. The best fits hint to a
neutrino mass $m_{\nu_\tau} \sim 5-7$ eV \cite{Primack}
which does appear naturally in our model. As for the
CDM itself, in our R parity conserving SUSY model it is
naturally provided by the lightest supersymmetric particle (LSP),
presumably a neutralino.

As we commented earlier, the neutrino mass hierarchy should be
qualitatively the same as that for the charged quarks and leptons.
However, the spread in the Yukawa coupling constants $h_F$
does not allow to put severe limits on the other neutrino masses.
For example, by taking $m_{\nu_\mu}/m_{\nu_\tau} \sim m_c/m_t$, as
is suggested by the first estimate in eq. (\ref{neutrino-hierarchy}),
one obtains $m_{\nu_\mu}\sim (2-5)\cdot 10^{-3}$. This range
corresponds to the Mikheyev-Smirnov-Wolfenstein
(MSW) solution of the solar neutrino problem
\cite{MSW} via $\nu_e\to \nu_\mu$ oscillations.
Alternatively, if we had to attempt an explanation
of the deficit of the atmospheric $\nu_\mu$ via $\nu_\mu\to \nu_e$
oscillations, then we would need $m_{\nu_\mu}\sim 0.1$ eV \cite{atm-new}
which is compatible with the second estimate in eq.
(\ref{neutrino-hierarchy}). Obviously the MSW explanation to the solar
neutrino deficit would not be viable in this latter case.
%via $\nu_e$--$\nu_\mu$ oscillations,
%and the mass ratios for the three neutrinos would indeed agree with the
%range suggested by the first estimate in (\ref{neutrino-hierarchy}).
%Clearly in this case the mass of the $\tau$ neutrino would be too
%small for this particle to play any major role as HDM.

%\vspace{2cm}

\section{Conclusions }

In this paper we have put forward the idea that natural
conservation of R parity in SUSY models can be guaranteed
in the presence of some suitable horizontal gauge symmetries.
We have shown how these symmetries
can indeed forbid all the dangerous terms in the superpotential,
which are trilinear in the fermion superfields, and how
an accidental $Z_2$ matter parity (equivalent to R parity)
then follows in a quite satisfactory way
only due to gauge invariance and to the field
content of the model.
On theoretical and phenomenological grounds, we have uniquely
identified $SU(4)_H$ as the only viable horizontal gauge group.
As a consequence, our scheme requires a fourth generation
of superfields in addition to the three known families.
Hence we have focused our analysis on a four generation SUSY model
based on the SM
vertical gauge group $SU(3)\times SU(2)\times U(1)$
and equipped with an $SU(4)_H$
anomaly free horizontal gauge symmetry.
We have discussed in some details the structure of the
fermion mass matrices arising in the model, as well as some
possible patterns for the breaking of the horizontal symmetry.
We have shown that the simplest symmetry breaking scheme which
ensures that all the horizontal modes acquire large masses,
can also lead to a particular form for the fermion mass matrices
which ensures that the masses for the fourth generation fermions
are naturally close to the electroweak scale.
A recent RG analysis of SUSY models with four generations
\cite{fourth-susy} does apply
straightforwardly to our case, and suggests that if
the hypothesis of unification of the vertical gauge group is
correct, then at least the
new leptons should be well in the reach of LEP II.
As regards the  masses of the first three families,
our model leads to a seesaw suppression of their magnitude
from the electroweak scale down to the observed values.
In particular,
this is achieved without the need of any tuning
for the Yukawa couplings, which can be assumed to be all $O(1)$
or close to the typical values of the gauge couplings.
By means of cosmological and astrophysical arguments,
we have managed to constrain rather precisely the scale $V_H$
at which the horizontal gauge symmetry is completely broken,
obtaining a very narrow window around $10^{11}\,$GeV.
Below this scale, our model is essentially the SSM with four generations.
In turn, the upper bound on the scale $V_H$ feeds back into
the neutrino mass matrix, implying a mass for the $\tau$-neutrino
not much lighter than a few eV.
A neutrino mass in this range will then give a sizeable contribution
to the present energy density of the Universe.
Thus, our model naturally provides cosmological HDM
in the form of $\nu_\tau$'s and, due to R parity conservation,
also CDM in the form of stable LSPs.
Since in our scheme conservation of R-parity is
ensured by the horizontal gauge symmetry independently
of the particular choice for the vertical gauge group,
it would be interesting to extend  the present analysis
to phenomenologically appealing GUT models,
such as $SU(5)$ or $E_6$, for which R-parity
conservation is not automatic.

\vspace{1cm}

{\large \bf Acknowledgements}

\vspace{0.4cm}
\noindent
It is a pleasure to thank Venya Berezinsky, Sasha Dolgov and Misha
Shaposhnikov for illuminating discussions and useful comments.
One of us (E.N.) would like to thank
I. Rothstein, R. Akhoury, M. Einhorn, G. Kane, R. Stuart, Y. Tomozawa,
M. Veltman, D. Williams, E. Yao, V. Zhakarov
and all the post-docs and students
of the Particle Theory Group of the University of Michigan,
for several useful discussions and
for the pleasant working atmosphere
during his stay at the University of Michigan.

                          \end{document}